\documentclass[prb,twocolumn,floatfix,superscriptaddress,showpacs,amssymb,
showkeys,amsmath]{revtex4}
\usepackage{graphicx}
\usepackage{amssymb}

\begin{document}
\title{Spin-orbit effects on two-electron states in nanowhisker double 
quantum dots}

\author{C.\ L.\ Romano }
\affiliation{Department of Physics ``J.\ J.\ Giambiagi",
University of Buenos Aires,
Ciudad Universitaria, Pabell\'on I,
C1428EHA Ciudad Aut\'onoma de Buenos Aires,
Argentina}
\author{P.\ I.\ Tamborenea}
\affiliation{Department of Physics ``J.\ J.\ Giambiagi",
University of Buenos Aires,
Ciudad Universitaria, Pabell\'on I,
C1428EHA Ciudad Aut\'onoma de Buenos Aires,
Argentina}
\author{S.\ E.\ Ulloa}
\affiliation{Department of Physics and Astronomy, and Nanoscale
and Quantum Phenomena Institute,
Ohio University, Athens, Ohio 45701-2979}
\date{\today}

\begin{abstract}
We investigate theoretically the combined effects of the
electron-electron and the Rashba spin-orbit interactions on two
electrons confined in quasi-one-dimensional AlInSb-based
double quantum dots.
We calculate the two-electron wave functions and explore the
interplay between these two interactions on the energy levels
and the spin of the states.
The energy spectrum as a function of an applied magnetic
field shows crossings and anticrossings between triplet and singlet
states, associated with level mixing induced by the spin-orbit coupling.
We find that the fields at which these crossings occur can be naturally 
controlled by the interdot barrier width, which controls the exchange  
integral in the structure.

\end{abstract}

\maketitle
\section{Introduction}
\label {sec:intro}

Substantial efforts have been devoted to understanding and
manipulating the electron spin aiming at potential applications in
semiconductor spintronics. \cite{pri,zut-fab-das,fab-etal} While
the spin-orbit interaction in quantum dots has been extensively
studied from the single-electron perspective, its combined effects
with the Coulomb interaction in few-particle systems has only
recently begun to be explored.
\cite{han-kou-pet-tar-van,wei-gol-ber,des-ull-mar,cha-pie}
Especially in nanowhiskers, Fasth {\it et al.} \cite{fas-etal}
measured the strength of the spin-orbit interaction in
two-electron InAs cylindrical dots (diameter $\sim50$~nm, length
$\sim120$~nm), and Pfund {\it et al.} \cite{pfu-etal} studied spin
relaxation in a similar system. 
Here we examine two-electrons in a
double quantum dot\cite{dyb-haw,abo-dyb-haw,xue-das,tam-met}
system with only one transverse quantum mode active. 
The one-electron spectrum of this thin nanowhisker semiconductor
structure has been treated by us previously, \cite{rom-ull-tam}
as well as the phonon-mediated spin-relaxation.
\cite{rom-tam-ull}

In this paper we calculate and analyze in detail the two-electron
states in such a system, taking into account the Rashba spin-orbit
coupling and the Coulomb interaction between the electrons. We pay
special attention to the degree of admixture of different
two-electron spin wave-functions, which will influence the
spin-flip transitions in this system.

The paper is organized as follows.
In Sec.\ \ref{sec:Hamiltonian} we introduce the effective
quasi-one-dimensional Hamiltonian of two interacting electrons in the
presence of the Rashba interaction and describe the method and
approximations used in the calculations.
In Sec.\ \ref{sec:results} we investigate the effects of the electron-electron
interaction on the energy levels with and without Rashba interaction and
their dependence on an applied magnetic field.
We monitor the mean value of the spin projection as a function of the
structural parameter that determines the strength of the Rashba
spin-orbit coupling.
In Sec.\ \ref{sec:conclusion} we provide some concluding remarks.

\section{Theoretical description}
\label{sec:Hamiltonian}

\begin{figure}[bth]
\includegraphics*[width=8cm]{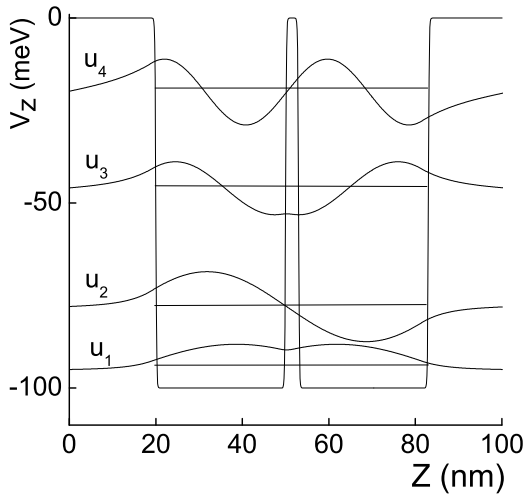}
\caption{The Al$_{0.1}$In$_{0.9}$Sb-InSb double-well confining
potential in the longitudinal direction of the quasi-one-dimensional nanowhisker quantum dots.  
The single-particle eigenfunctions and energies are also shown. }
\label{fig:potential}
\end{figure}

We investigate the problem of two interacting electrons in a
quasi-one-dimensional double quantum dot structure in the presence
of the structural or Rashba spin-orbit interaction.
We study two identical 30~nm wide dots, separated by an interdot
barrier, 3~nm or 5~nm wide.
In our calculations we consider an
Al$_{0.1}$In$_{0.9}$Sb-InSb structure which has a potential energy
depth of $100 \, \mbox{meV}$. In Fig.\ \ref{fig:potential} we show
the confining potential in the longitudinal direction, $V_z(z)$,
and the eigen-functions $u_n(z)$ of the single-particle
Hamiltonian $H^0 = \frac{p_z^2}{2m^*} + V_z(z)$, displaced
vertically according to their corresponding energy levels, $E_n$.

The nanowhisker where the double dot structure is defined is assumed to be so thin ($\simeq 2$ nm) 
that only the lowest transverse mode is active.  
As such, the effective one-dimensional Hamiltonian of two interacting
electrons with Rashba interaction, in the absence of a magnetic
field, can be written as \cite{rom-ull-tam}
\begin{equation}
H \, = \, H^0_1 + H^0_2 +  H_{1dR} + V_{int},
\label{eq:H1d}
\end{equation}
where $H^0_i = \frac{p_{z,i}^2}{2m^*} + V_z(z_i)$,
$m^*$ is the conduction-band effective mass,
$z_1$ and $z_2$ are the $z$-coordinates of the two electrons,
and $p_{z,1}$ and $p_{z,2}$ are the $z$-components of their
linear momentum.
$H_{1dR}$ and $V_{int}$ are
the Rashba spin-orbit coupling
and
the electron-electron interaction potential,
respectively.
The Rashba spin-orbit coupling in the quasi-one-dimensional structure
considered here is given by \cite{rom-ull-tam}
\begin{equation}
H_{1dR} = \sum^2_{i=1}
          \frac{\gamma_R}{\hbar}
          \left\langle \frac{\partial V_x}{\partial x} \right\rangle
          p_{z,i}
          (\sigma_{x_i}-\sigma_{y_i}),
\end{equation}
where $\gamma_R = 500 \, \mbox{\AA}^2$, \cite{vos-lee-tre}
$\sigma_{x_i,y_i}$ are Pauli matrices, and
$\left\langle\frac{\partial V_x}{\partial x}\right\rangle$
is the Rashba field parameter, where the mean value is taken over the ground 
state  $\Phi$ of the laterally-confining potential $V_x=V_y$
($V_x$ is assumed to be the same as $V_y$ for simplicity).
The electron-electron interaction is given by \cite{tam-met}
\begin{eqnarray}
    V_{int}(|z_2-z_1|) =
      \int \, dx_1 dx_2 dy_1 dy_2  \nonumber \\
     \frac{e^2 \Phi(x_1)^2 \Phi(x_2)^2 \Phi(y_1)^2 \Phi(y_2)^2}
           {\epsilon \sqrt{(x_1-x_2)^2+(y_1-y_2)^2+(z_1-z_2 )^2}},
\label{eq:Vint}
\end{eqnarray}
where $\textbf{r}_i = (x_i,y_i,z_i)$, $i=1,2$, are the electron
positions, and $\epsilon$ is the dielectric constant of the
material (16.8 for InSb).  
In Eq.\ (\ref{eq:Vint}) we approximate $\Phi$ by the ground state (spatial 
extent $\sim$ 2 nm) of a harmonic oscillator potential defined over the 
cross section of the whisker.

As a basis set for the two-electron Hilbert space we take
\begin{eqnarray}
  \varphi_1 &=& u_1(z_1) u_1(z_2) \,
                |0,0 \rangle, \nonumber\\
  \varphi_2 &=& \frac{1}{\sqrt{2}}
                \left(u_1(z_1) u_2(z_2) + u_2(z_1) u_1(z_2) \right)
                |0,0 \rangle, \nonumber\\
  \varphi_3 &=& u_2(z_1) u_2(z_2)
                |0,0 \rangle, \nonumber\\
  \varphi_4 &=& \frac{1}{\sqrt{2}}
                \left(u_1(z_1) u_2(z_2) - u_2(z_1) u_1(z_2) \right)
                |1,1 \rangle,                              \nonumber\\
  \varphi_5 &=& \frac{1}{\sqrt{2}}
                \left(u_1(z_1) u_2(z_2) - u_2(z_1) u_1(z_2) \right)
                |1,-1 \rangle,                             \nonumber\\
  \varphi_6 &=& \frac{1}{\sqrt{2}}
                \left(u_1(z_1) u_2(z_2) - u_2(z_1) u_1(z_2) \right)
                |1,0 \rangle.
\end{eqnarray}
The two-particle spin wave functions are the usual singlet
$|S\rangle = |S=0,m_S=0 \rangle$
and triplet states
$\{|T^+\rangle = |1,1 \rangle,
|T^0\rangle = |1,0 \rangle,
|T^-\rangle = |1,-1 \rangle\}$.
This truncated basis set $\{\varphi_i, i=1,6\}$ takes into
account only the two lowest eigenstates of $H^0$, $u_1(z)$ and
$u_2(z)$ (see Fig.\ \ref{fig:potential}).
This truncated basis simplifies the analysis and presentation
without leaving out essential physics.
We have checked that the truncation introduces only small quantitative
differences, of the order of 1 percent, in the results.

In order to solve the eigenvalue problem of the full two-electron
Hamiltonian given in Eq.\ (\ref{eq:H1d}), we expand the
two-electron wave functions
\begin{equation}
\psi_i = \sum_{j=1}^6 a_{ij} \, \varphi_j,
\end{equation}
where $i = 1,..,6$, and determine the coefficients $a_{ij}$ by numerical
diagonalization.  
It is clear that the Rashba coupling mixes states with different spin wave 
functions, although the mixing depends strongly in structure parameters 
and applied magnetic field, as we will see in the next section.

\begin{figure}[bth]
\includegraphics*[width=8cm]{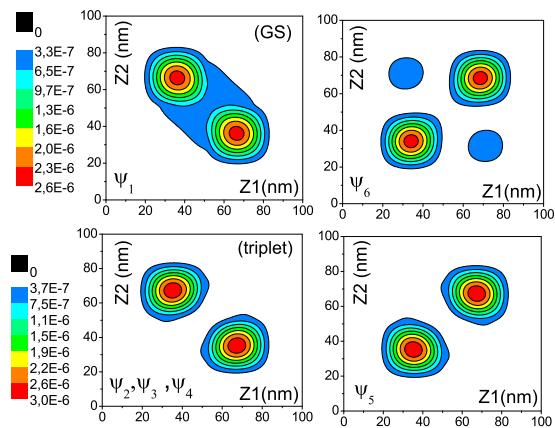}
\caption{(color online) Probability density of two electrons in the double
quantum dot with Coulomb interaction but without Rashba coupling.
$\psi_{1}$: ground state (GS), $\psi_{i=2,...,6}$ excited states.  
Notice $\psi_5$ and $\psi_6$ describe states with both electrons in the 
{\em same} dot, while the others can be seen more as having one electron 
in each dot.}
\label{fig:modulocuadrado}
\end{figure}

\section{Results}
\label{sec:results}

\begin{figure}[hbt]
\includegraphics*[width=9cm]{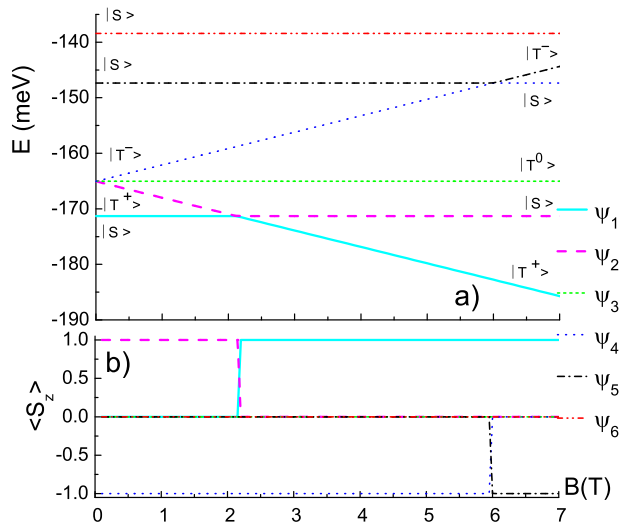}
\caption{(color online) (a) Energy levels and (b) mean value of 
$S_z=S_{1,z}+S_{2,z}$
         {\it vs.} applied magnetic field for the two-electron eigenstates
         with Coulomb interaction and without Rashba coupling.
         Barrier width: 3~nm. $|S\rangle =$ singlet state,
         $|T^{\pm,0}\rangle =$ triplet states. }
\label{fig:E_Sz_sin_Rashba}
\end{figure}

We solve the eigenvalue problem of the two-electron Hamiltonian
$H$ given in Eq.\ (\ref{eq:H1d}) with the goal of
understanding the interplay between the electron-electron Coulomb
interaction and the single-particle Rashba coupling.
In order to gain some initial insight into the nature of the
two-electron states, in Fig.\ \ref{fig:modulocuadrado} we show the
probability density of the six eigenstates for the two electrons
in a double-dot structure with a 3~nm barrier {\em without} Rashba
coupling.
As expected, in the ground state the two electrons are essentially localized
in different dots, due to their mutual Coulomb repulsion, and on ``bonding" 
single-particle states with a non-zero amplitude in the interdot barrier 
region  
In contrast, states $\psi_2$ to $\psi_4$ have electrons localized in each 
dot too, but with an ``antibonding" orbital with zero amplitude in the 
central barrier.  
Notice also that the two higher energy states, $\psi_5$ and $\psi_6$, 
correspond to singlet states with two electrons in each dot.
Furthermore, in order to fully characterize the two-electron
system in what could be realistic experimental situations we
introduce a magnetic field along the $z$-direction.
The field is chosen small enough (and the whisker so thin) that
the $x$-$y$ orbital wave functions are not perturbed significantly by it.
Thus, the field contributes only a Zeeman term to the Hamiltonian:
\begin{equation}
H_Z = \frac{g_0 \mu_B B}{\hbar} S_z,
\end{equation}
where $\mu_B$ is the Bohr magneton, $g_0$ is the Land\'e factor
($g_0=-51$ for InSb),
and
$S_z=S_{1,z}+S_{2,z}$ is the $z$-component of the total spin operator.
In Fig.\ \ref{fig:E_Sz_sin_Rashba} we plot the energy and expectation
value of $S_z$, for all the two-electron eigenstates {\it vs.} the
applied magnetic field {\it without} Rashba interaction,
in the double dot with a 3~nm barrier.

Let us point out some basic features of the results without Rashba
coupling seen in Fig.\ \ref{fig:E_Sz_sin_Rashba}.
First notice that since the spatial wave functions do not depend on the 
magnetic field and only the Zeeman energy changes with $B$, this
explains the linear field dependence of the energies.
At zero magnetic field, the ground state $\psi_1$ is a singlet
($|0,0\rangle$, $S_z=0$), but around $B \approx 2.2 \, \mbox{T}$
there is a level crossing and its spin part becomes $|1,1\rangle$
($S_z=1$).
This change occurs due to the competition between the Zeeman energy,
on one hand, and the Coulomb interaction and $E_2 - E_1$ on the other.
At the crossing, the first excited state
$\psi_2$ goes, naturally, from $S_z=1$ to $S_z=0$.
$\psi_2$, $\psi_3$ and $\psi_4$ are degenerate at $B=0$, but
having $S_z=1,0,-1$, respectively, their degeneracy is broken for
non-zero $B$.
$E_4$ and $E_5$ have a level crossing at $B \approx 6\, \mbox{T}$,
and, finally, $E_5$ and $E_6$ cross at $B > 7 \, \mbox{T}$ (not shown).
There are no further crossings at higher magnetic fields.

\begin{figure}[bht]
\includegraphics*[width=9cm]{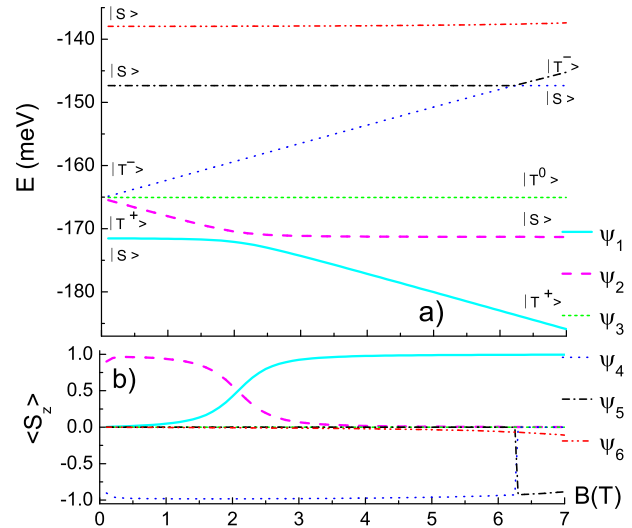}
\caption{(color online) (a) Energy levels and (b) mean value of
$S_z=S_{1,z}+S_{2,z}$
        {\it versus} applied magnetic field for the two-electron eigenstates
         including Coulomb and Rashba interactions.
         The strength of the Rashba coupling is given by
         $\left\langle \frac{\partial V_x}{\partial x} \right\rangle
       = 1 \, \mbox{meV/\AA}$. Barrier width: 3~nm. $|S\rangle =$ singlet state,
         $|T^{\pm,0}\rangle =$ triplet state.}
\label{fig:E_Sz_con_Rashba}
\end{figure}

\begin{figure}[thb]
\includegraphics*[width=9cm]{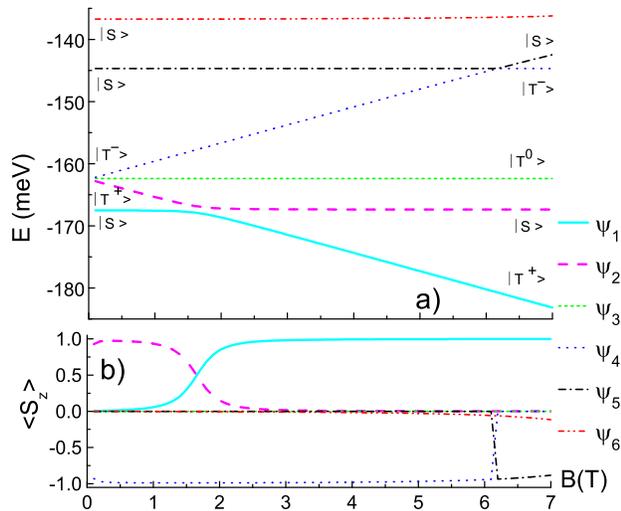}
\caption{ (color online) (a) Energy levels and (b) mean value of
$S_z=S_{1,z}+S_{2,z}$
         {\it versus} applied magnetic field for the two-electron eigenstates
         including Coulomb and Rashba interactions.
         The strength of the Rashba coupling is given by
         $\left\langle \frac{\partial V_x}{\partial x} \right\rangle
      = 1 \, \mbox{meV/\AA}$. Barrier width: 5~nm. $|S\rangle =$ singlet state,
         $|T^{\pm,0}\rangle =$ triplet state. (color online)}
\label{fig:ESzconRashbab10}
\end{figure}

We next include the Rashba interaction with a coupling parameter
$\left\langle \frac{\partial V_x}{\partial x}\right\rangle=1 \,
\mbox{meV/\AA}$.
The results for the energy levels and mean value of $S_z$ are shown
in Fig.\ \ref{fig:E_Sz_con_Rashba}. 
The main differences with Fig.\ \ref{fig:E_Sz_sin_Rashba} are as follows: \\
(i) With Rashba interaction, at $B=0$, $\psi_2$ and $\psi_4$ are
not spin eigenstates anymore, as can be seen in
Fig.\ \ref{fig:E_Sz_con_Rashba}(b).
However, we still label the states as if they were pure spin states,
as one spin state dominates the admixture (far from the avoided
crossings discussed next). \\
(ii) Two of the level crossings in Fig.\ \ref{fig:E_Sz_sin_Rashba} 
($E_1$ with $E_2$ at $B \approx 2.2 \, \mbox{T}$,
and $E_5$ with $E_6$ at $B > 7 \, \mbox{T}$) become avoided crossings here, 
as the pair of states involved are coupled by the Rashba interaction.
The other crossing, between $E_4$ and $E_5$ at $B \approx 6\, \mbox{T}$,
is shifted slightly upward due to the effect of the Rashba interaction
on each individual level,
but it does not become avoided because the levels are not coupled to
each other through the Rashba interaction.  
This lack of mixing arises from the different spatial symmetry of the states 
and the strong Coulomb interaction.  
The symmetry under space reversal (odd {\em vs.} even) prevents the mixing 
of a state with double dot occupancy (singlet $\psi_5$) and a state where 
each dot has one electron ($\psi_4$), where each state has opposite 
space-reversal symmetry.
Notice that the width of the avoided crossings is determined mainly
by the strength of the spin-orbit coupling, and therefore it can be
adjusted with a transverse electric field (gate voltage).

In Fig.\ \ref{fig:ESzconRashbab10} we present the energy levels and
the mean value of $S_z$ as functions of the magnetic field for a structure
with a 5~nm interdot barrier.
Notice that, compared to the case of a 3~nm interdot barrier
(Fig.\ \ref{fig:E_Sz_con_Rashba}),
crossings and anti-crossings shift to lower values of
the magnetic field.
This happens because the wider barrier decreases the bonding-antibonding gap,
 the wave function overlap and the associated exchange integral that 
determines the singlet-triple separation at zero field.  
As the $S$-$T$ gap is smaller, it is more easily overcome by the Zeeman 
energy.  
For an 8~nm barrier, the bonding-antibonding gap is only slightly over half 
of the value for the 3~nm barrier, and the singlet-triplet crossing occurs 
at a field of 1.2 T, for example.  

\begin{figure}[tbh]
\includegraphics*[width=9cm]{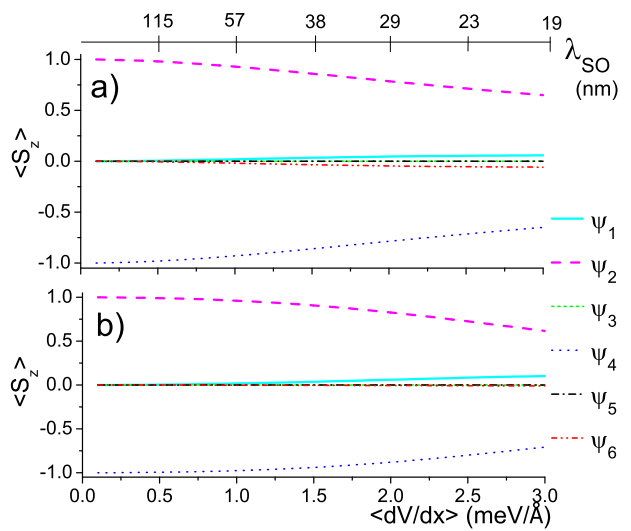}
\caption{(color online) Mean value of $S_z=S_{1,z}+S_{2,z}$ as a function of 
the Rashba parameter
         $\left\langle \frac{\partial V_x}{\partial x}\right\rangle$
         for the six two-electron eigenstates. $B=0.5 \,\mbox{T}$.
         Barrier width: 3~nm.
         (a) Without Coulomb interaction;
         (b) With Coulomb interaction. $\lambda_{SO}$ is the spin-orbit
         length.
        }
\label{fig:Sz_vs_dVdx}
\end{figure}

In Fig.\ \ref{fig:Sz_vs_dVdx} we present the mean value of $S_z$ as a
function of the Rashba parameter
$\left\langle \frac{\partial V_x}{\partial x}\right\rangle$
for all the states at a given magnetic field $B=0.5\,\mbox{T}$
(using a barrier width of 3~nm).
In this figure, we compare two cases, (a) without and (b) with Coulomb
interaction, in order to exhibit better the role of spin-orbit coupling in 
the spin mixing.
As expected, without Rashba coupling, {\em i.e}.\@ with
$\left\langle \frac{\partial V_x}{\partial x}\right\rangle$=0, the spin 
projection of each of these states naturally
$\langle S_z \rangle$ takes the exact values 1, 0, and $-1$,
as seen in both Figs.\ \ref{fig:Sz_vs_dVdx}(a) and (b).
An important difference between (a) and (b) is
that there is a symmetry around $S_z=0$ when the Coulomb
interaction is absent.
The ground state in Fig.\ \ref{fig:Sz_vs_dVdx}(a) (thick cyan solid line)
starts with $S_z=0$ and at a certain (typically large) value of the Rashba 
parameter reaches a maximum.
On the other hand, in Fig.\ \ref{fig:Sz_vs_dVdx}(b), this symmetry about the zero value 
is lost due to the different mixing of two-particle orbitals ($\{\varphi_i\}$) in higher- 
and lower-lying states produced by the Coulomb interaction. 
Figure \ref{fig:Sz_vs_dVdx} includes an axis (top) in terms of the spin-orbit length 
($\lambda_{SO} = \hbar^2 / m^*\gamma_R 
\left\langle 
             \frac{\partial V_x}{\partial x}
\right\rangle$),\cite{bul-los} 
which is inversely proportional to the Rashba parameter. 
This length parameter helps visualize the strength of the Rashba
coupling in comparison to the characteristic dimensions of the structure. 
It is interesting to point out that spin mixing is first noticeable 
when $\lambda_{SO} \simeq 60$ nm, the size of the two-well system.

With a 5~nm barrier width, we find that the ground state has a mean
value of $S_z$ roughly equal to zero up to a Rashba constant of
2.2~\mbox{meV/\AA} in the presence of the Coulomb interaction, unlike the 
case in Fig.\ \ref{fig:Sz_vs_dVdx}, where the ground state is clearly 
mixed for Rashba constant beyond 1.5 meV/\AA.

\section{Conclusions}
\label{sec:conclusion}

In this paper, we have investigated the effects of the Coulomb
electron-electron interaction on the states, energy levels, and 
\textit{z}-projection of the spin of two electrons confined in a 
quasi-1D double-quantum-dot nanowhisker in the presence of Rashba 
spin-orbit interaction.
As a function of a magnetic field in the longitudinal direction,
some energy-level crossings become avoided crossings when the Rashba
spin-orbit is turned on.
The width of the avoided crossings can be controlled with a lateral gate 
voltage via the intensity of the Rashba parameter
$\left\langle \frac{\partial V_x}{\partial x}\right\rangle$.
The positions of these crossings and avoided crossings as functions of the 
magnetic field can be selected by changing the width of the interdot barrier.
Finally, we found that the Coulomb interaction reduces the spin mixing in the 
ground state, as well as in the excited states, and displayed this reduction
as a function of the Rashba spin-orbit length.

\begin{acknowledgments}
We acknowledge useful discussions with C.\ Destefani and L.\ Meza-Montes, 
as well as support from CONICET (PIP-5851), UBACyT (X179), and NSF 
(WMN grant 0710581). 
PIT is a researcher of CONICET.
\end{acknowledgments}

\end{document}